\documentclass[12pt]{article}
\usepackage{epsfig}
\usepackage{amsfonts}
\usepackage{latexsym}
\usepackage{amsmath,amssymb}
\usepackage{mathrsfs}
\usepackage{hyperref}
\usepackage{setspace}
\usepackage{color}
\usepackage{bm}
\usepackage{slashed}
\numberwithin{equation}{section}

\begin{document}

\begin{titlepage}
\unitlength = 1mm
%\today 
\begin{flushright}
YITP-18-03
\end{flushright}

\vskip 1cm
\begin{center}

{\Large {\textsc{\textbf{Quantum entanglement in de Sitter space \\
        with a wall, and the decoherence of bubble universes}}}}

\vspace{1.8cm}
Andreas Albrecht$^*$, Sugumi Kanno$^{\flat\,\natural}$ and Misao Sasaki$^{\dag\,\ddag}$

\vspace{1cm}

\shortstack[l]
{\it $^*$ Center for Quantum Mathematics and Physics and Department of Physics,\\ 
~~\it University of California Davis, Davis, California, 95616, USA\\
\it $^\flat$ Department of Theoretical Physics and History of Science,\\
~~\it University of the Basque Country, Bilbao, 48080, Spain\\
\it $^\natural$ IKERBASQUE, Basque Foundation for Science, 
Maria Diaz de Haro 3,
Bilbao, 48013, Spain\\
\it $^\dag$ Center for Gravitational Physics,
Yukawa Institute for Theoretical Physics,\\
~~\it Kyoto University, Kyoto 606-8502, Japan\\
\it $^\ddag$ International Research Unit of Advanced Future Studies,
 Kyoto University, Japan
}

\vskip 1.5cm

\begin{abstract}
\baselineskip=6mm
We study the effect of a bubble wall on the entanglement entropy of a free massive 
scalar field between two causally disconnected open charts in de Sitter space. 
We assume there is a delta-functional wall between the open charts. 
This can be thought of as a model of pair creation of
bubble universes in de Sitter space. We first derive the Euclidean vacuum mode 
functions of the scalar field in the presence of the wall in the coordinates 
that respect the open charts. We then derive the Bogoliubov transformation 
between the Euclidean vacuum and the open chart vacua that makes the 
reduced density matrix diagonal. We find that larger walls lead to less entanglement. Our result may be regarded as evidence of decoherence of bubble universes from each other. We also note an interesting relationship between our results and discussions of the black hole firewall problem. 
\end{abstract}

\vspace{1.0cm}

\end{center}
\end{titlepage}

\pagestyle{plain}
\setcounter{page}{1}
\newcounter{bean}
\baselineskip18pt

\setcounter{tocdepth}{2}

\tableofcontents

\section{Introduction}

Quantum entanglement has fascinated many physicist because of its counterintuitive nature.  Quantum entanglement makes it possible to know everything about a system composed of two subsystems (in a pure state) but know nothing at all about the subsystems (in the case of maximal entanglement).
After Aspect et al. succeeded in showing experimental evidence of the quantum nature of entanglement by measuring correlations of linear polarizations of pairs of photons~\cite{Aspect:1981zz, Aspect:1982fx}, much attention has been paid to this genuine quantum property in various research areas including quantum information theory, quantum communication, quantum cryptography, quantum teleportation and quantum computation.

Quantum entanglement should play an important role in cosmology. In de Sitter space where the universe expands exponentially, any two mutually separated regions eventually become causally disconnected. This is most conveniently described by spanning open universe coordinates on two open charts in de Sitter space. The positive frequency mode functions of a free massive scalar field for the Euclidean vacuum (the Bunch-Davies vacuum) that have support on both regions were derived in~\cite{Sasaki:1994yt}. Using them, quantum entanglement between two causally disconnected regions in de Sitter space was first studied by Maldacena and Pimentel~\cite{Maldacena:2012xp}. They showed that the entanglement entropy, which is a measure of quantum entanglement, of a free massive scalar field between two disconnected open charts is non-vanishing. Motivated by this, the entanglement entropy of $\alpha$-vacua~\cite{Kanno:2014lma, Iizuka:2014rua}, that of the Dirac field~\cite{Kanno:2016qcc} and axion field were examined~\cite{Choudhury:2017bou, Choudhury:2017qyl}. The spectrum of cosmological fluctuation was also studied in~\cite{Kanno:2014ifa, Dimitrakopoulos:2015yva}. Quantum entanglement is also of considerable interest in the context of the proposed ``entanglement -- geometry correspondence''(e.g.~\cite{Ryu:2006bv, Hubeny:2007xt}). 

One of the cornerstones of inflationary cosmology is that primordial density fluctuations have a quantum mechanical origin. Inflation leads to an ``initial state'' of the universe following inflation which is highly entangled. This invites the question of whether compelling observational evidence for the entangled nature of the initial density fluctuations can be found. Several studies have been made on quantifying the initial state entanglement by using some measure of entanglement such as the Bell inequality~\cite{Campo:2005sv, Campo:2005qn, Maldacena:2015bha, Martin:2016tbd, Choudhury:2016cso, Kanno:2017dci, Martin:2017zxs}, entanglement negativity~\cite{Nambu:2008my, Kanno:2014bma, Matsumura:2017swh} and quantum discord~\cite{Martin:2015qta, Kanno:2016gas}. There have also been several attempts to find some observational signatures on the CMB when the initial state is a non-Bunch-Davies vacuum due to entanglement between two scalar fields~\cite{Albrecht:2014aga, Kanno:2015ewa}, between two universes~\cite{Kanno:2015lja}, and due to scalar-tensor entanglement~\cite{Collins:2016ahj, Bolis:2016vas}

In this paper we extend the calculation of Maldacena and Pimentel~\cite{Maldacena:2012xp} to the case where a bubble wall is present between the two open charts. The modes of the scalar field are changed by the presence of the wall, which in turn changes the entanglement entropy between the two regions. We find that for sufficiently large walls, the entanglement entropy approaches zero. Our technical results may prove useful in several of the areas discussed above.  Here we focus on the possible implications for the decoherence of bubble universes.

The paper is organized as follows. In section 2, we review the method developed by Maldacena and Pimentel with some comments relevant to the calculation of the entanglement entropy with a bubble wall. In section 3, we introduce the bubble wall in the system and construct the positive frequency mode functions for the Bunch-Davies vacuum. We then compute the entanglement entropy and logarithmic negativity. Finally we summarize our result and discuss the implications in section 4.

\section{Entanglement entropy in de Sitter space}
\noindent

Recently, Maldacena and Pimentel studied quantum entanglement between two causally disconnected regions in de Sitter space in~\cite{Maldacena:2012xp}. They showed that the entanglement entropy of a free massive scalar field between two disconnected open charts is non-vanishing. In this section, we review their result.

\subsection{Mode functions in the open chart}

\begin{figure}[t]
\vspace{-2cm}
\includegraphics[height=8cm]{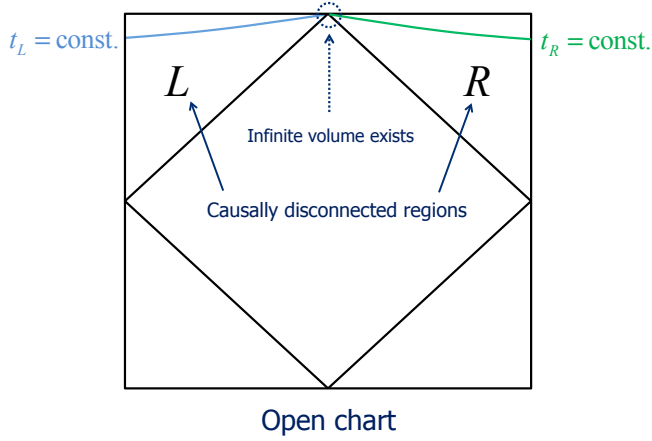}\centering
\vspace{-1.2cm}
\caption{The Penrose diagram of the de Sitter space is shown. 
$L$ and $R$ are the two causally disconnected regions
described by the open charts. A late-time spatial hypersurface 
in each region is depicted.}
\label{fig1}
\end{figure}

We consider a free scalar field $\phi$ with mass $m$ in de Sitter space 
represented by the metric $g_{\mu\nu}$.  The action is given by
\begin{eqnarray}
S=\int d^4 x\sqrt{-g}\left[\,-\frac{1}{2}\,g^{\mu\nu}
\partial_\mu\phi\,\partial_\nu \phi
-\frac{m^2}{2}\phi^2\,\right]\,.
\label{action}
\end{eqnarray} 
The metric in each $R$ and $L$ region of open charts in de Sitter space 
(see  Figure~\,\ref{fig1}) can be obtained by analytic continuation from 
the Euclidean metric,
\begin{align}
ds^2_E=H^{-2}\left[d\tau^2+\cos^2\tau\left(d\rho^2+\sin^2\rho\,d\Omega^2\right)
\right]\,,
\label{Emetric}
\end{align}
 and expressed, respectively, as
\begin{eqnarray}
ds^2_R&=&H^{-2}\left[-dt^2_R+\sinh^2t_R\left(dr^2_R+\sinh^2r_R\,d\Omega^2\right)
\right]\,,\nonumber\\
ds^2_L&=&H^{-2}\left[-dt^2_L+\sinh^2t_L\left(dr^2_L+\sinh^2r_L\,d\Omega^2\right)
\right]\,,
\end{eqnarray}
where $H^{-1}$ is the Hubble radius and $d\Omega^2$ is the metric on the two-sphere. 
Note that the region $R$ and $L$ covered by the coordinates $(t_L, r_L)$ and $(t_R, r_R)$ respectively are the two causally  disconnected open charts of de Sitter 
space\footnote{The point between $R$ and $L$ regions is a part of the timelike 
infinity where infinite volume exists.}. 

The solutions of the Klein-Gordon equation are expressed as
\begin{eqnarray}
u_{\sigma p\ell m}(t,r,\Omega)\sim\frac{H}{\sinh t}\,
\chi_{p,\sigma}(t)\,Y_{p\ell m} (r,\Omega)\,,\qquad
-{\rm\bf L^2}Y_{p\ell m}=\left(1+p^2\right)Y_{p\ell m}\,,
\end{eqnarray}
where $(t,r)=(t_R,r_R)$ or $(t_L,r_L)$ and $Y_{p\ell m}$ are harmonic functions on the three-dimensional hyperbolic space. The eigenvalues $p$ normalized by $H$ take positive real values. The positive frequency mode functions corresponding to the Euclidean vacuum (the Bunch-Davies vacuum) that are supported both on the $R$ and $L$ regions are derived by Sasaki, Tanaka and Yamamoto in~\cite{Sasaki:1994yt}:
\begin{eqnarray}
\chi_{p,\sigma}(t)=\left\{
\begin{array}{l}
\frac{e^{\pi p}-i\sigma e^{-i\pi\nu}}{\Gamma(\nu+ip+\frac{1}{2})}P_{\nu-\frac{1}{2}}^{ip}(\cosh t_R)
-\frac{e^{-\pi p}-i\sigma e^{-i\pi\nu}}{\Gamma(\nu-ip+\frac{1}{2})}P_{\nu-\frac{1}{2}}^{-ip}(\cosh t_R)
\,,\\
\\
\frac{\sigma e^{\pi p}-i\,e^{-i\pi\nu}}{\Gamma(\nu+ip+\frac{1}{2})}P_{\nu-\frac{1}{2}}^{ip}(\cosh t_L)
-\frac{\sigma e^{-\pi p}-i\,e^{-i\pi\nu}}{\Gamma(\nu-ip+\frac{1}{2})}P_{\nu-\frac{1}{2}}^{-ip}(\cosh t_L)
\,,
\label{solutions}
\end{array}
\right.
\end{eqnarray}
where $P^{\pm ip}_{\nu-\frac{1}{2}}$ are the associated Legendre functions
and the index $\sigma$ takes the values $\pm 1$ which distinguishes two 
independent solutions for each region, and $\nu$ is a mass parameter
\begin{eqnarray}
\nu=\sqrt{\frac{9}{4}-\frac{m^2}{H^2}}\,.
\end{eqnarray}
Here and below in the text, we focus on the case $m^2/H^2<9/4$
to save space and make discussion clear.
The extension to the case $m^2/H^2>9/4$ is straightforward,
and the result we present will include both mass ranges.

Note that $\nu=1/2$ ($m^2=2H^2$) is equivalent to a conformally coupled 
massless scalar. The minimally coupled massless limit is $\nu=3/2$.
For $1/2<\nu<3/2$, it is known that there exists a supercurvature mode
$p=ik$ where $0<k<1$, which may be regarded as a bound-state
mode. The role of supercurvature modes in the quantum entanglement is not clear.
In~\cite{Maldacena:2012xp}, it is conjectured that they won't contribute.
In the body of this paper we simply ignore them. An analysis in the case of
a conformal scalar in the presence of a bubble wall is given in the
Appendix~\ref{app:a}. It turns out that a bubble wall can make the effective potential
deep and allow a supercurvature mode to exist. In fact,
we find that the eigenvalue $k$ can exceed unity
and become arbitrarily large as the effective potential becomes deeper,
and as a result the contribution of the supercurvature mode
in the vacuum spectrum in each open chart is more important\footnote{See Eq.~(3.10) in~\cite{Yamamoto:1996qq}}. 

Going back to the solutions in Eq.~(\ref{solutions}),
the Klein-Gordon normalization fixes the normalization factor as
\begin{eqnarray}
N_{p}=\frac{4\sinh\pi p\,\sqrt{\cosh\pi p-\sigma\sin\pi\nu}}{\sqrt{\pi}\,|\Gamma(\nu+ip+\frac{1}{2})|}\,.
\label{norm}
\end{eqnarray}
Since they form a complete orthonormal set of modes, the field can be
 expanded in terms of the creation and annihilation operators,
\begin{eqnarray}
\hat\phi(t,r,\Omega) &=& \frac{H}{\sinh t}\int dp \sum_{\sigma,\ell,m} 
\left[\,a_{\sigma p\ell m}\,\chi_{p,\sigma}(t)
+a_{\sigma p\ell -m}^\dagger\,\chi^*_{p,\sigma}(t)\,\right]Y_{p\ell m}(r,\Omega)
\nonumber\\
&=&\frac{H}{\sinh t}\int dp \sum_{\ell,m}\phi_{p\ell m}(t)Y_{p\ell m}(r,\Omega)
\,,
\end{eqnarray}
where $Y_{p\ell m}^*=Y_{p\ell -m}$,
$[a_{\sigma p\ell m},a_{\sigma' p'\ell' m'}^\dag]=\delta(p-p')\delta_{\sigma,\sigma'}\delta_{\ell,\ell'}\delta_{m,m'}$, and 
$a_{\sigma p\ell m}$ annihilates the Bunch-Davies vacuum,
$a_{\sigma p\ell m}|0\rangle_{\rm BD}=0$,
and we introduced a Fourier mode field operator,
\begin{eqnarray}
\phi_{p\ell m}(t)\equiv
\sum_\sigma\left[\,a_{\sigma p\ell m}\,\chi_{p,\sigma}(t)
+a_{\sigma p\ell -m}^\dagger\,\chi^*_{p,\sigma}(t)\right]\,.
\label{phi1}
\end{eqnarray}

For convenience, we write the mode functions and the associated Legendre 
functions of the $R$ and $L$ regions in a simple form
 $\chi_{p,\sigma}(t)\equiv\chi^{\sigma}$,
$P_{\nu-1/2}^{ip}(\cosh t_{R,L})\equiv P^{R, L}$,
$P_{\nu-1/2}^{-ip}(\cosh t_{R,L})\equiv P^{R*, L*}$.
 Also below we omit the indices $p$, $\ell$, $m$ of $\phi_{p\ell m}$, 
$a_{\sigma p\ell m}$ and $a_{\sigma p\ell -m}^\dag$ for simplicity.
For example, $a_\sigma=a_{\sigma p\ell m}$ and $a_\sigma^\dag=a_{\sigma p\ell -m}$ 
 unless there may be any confusion.\footnote{It may be noted that this 
abbreviation implies
$(a_\sigma)^\dag=a_{\sigma p\ell m}^\dag\neq a_\sigma^\dag=a_{\sigma p\ell -m}^\dag$.
But since this is a small technical problem that can be easily solved by 
doubling the degrees of freedom, below we assume
$(a_\sigma)^\dag=a_\sigma^\dag$.}

\subsection{Bogoliubov transformations and entangled states}

Next we consider the positive frequency
mode functions for the $R$ or $L$ vacuum that 
have support only on the $R$ or $L$ region, respectively. 
They are given by
\begin{eqnarray}
\varphi^q=\left\{
\begin{array}{ll}
\tilde{N}_p^{-1}P^q~&\mbox{in region}~q\,,
\\
0~ &\mbox{in the opposite region}\,,
\end{array}
\right.
\quad\tilde{N}_p=\frac{\sqrt{2p}}{|\Gamma(1+ip)|}\,,
\label{varphi}
\end{eqnarray}
where $q=(R, L)$. As the Fourier mode field operator (\ref{phi1}) should 
be the same under this change of mode functions, we have
\begin{eqnarray}
\phi(t)=a_\sigma\,\chi^\sigma+a_\sigma^\dag\,\chi^\sigma{}^*
=b_q\,\varphi^q+b_q^\dag\,\varphi^q{}^*\,,
\label{fo}
\end{eqnarray}
where we have introduce the new creation and annihilation operators ($b_q,b_q^\dag$)
such that $b_q|0\rangle_{q}=0$.
The operators $(a_\sigma,a_\sigma^\dag)$ and $(b_q,b_q^\dag)$
are related by a Bogoliubov transformation. 
The Bunch-Davies vacuum may be constructed from the states
over $|0\rangle_{q}$ as
\begin{eqnarray}
|0\rangle_{\rm BD}\propto\exp\left(\frac{1}{2}\sum_{i,j=R,L}
m_{ij}\,b_i^\dagger\, b_j^\dagger\right) |0\rangle_R|0\rangle_L\,,
\label{bogoliubov1}
\end{eqnarray}
where $m_{ij}$ is a symmetric matrix.
The condition $a_\sigma|0\rangle_{\rm BD}=0$ determines $m_{ij}$:
\begin{eqnarray}
m_{ij}=e^{i\theta}\frac{\sqrt{2}\,e^{-p\pi}}{\sqrt{\cosh 2\pi p+\cos 2\pi\nu}}
\left(
\begin{array}{cc}
\cos \pi\nu & i\sinh p\pi \vspace{1mm}\\
i\sinh p\pi & \cos \pi\nu \\
\end{array}
\right)\,,
\label{mij}
\end{eqnarray}
where $e^{i\theta}$ contains all unimportant phase factors for $\nu^2>0$. 
This is an entangled state of the ${\cal H}_R\otimes{\cal H}_L$ Hilbert space.

The density matrix $\rho=|0\rangle_{\rm BD}\,{}_{\rm BD}\langle0|$ is not diagonal 
in the $|0\rangle_R|0\rangle_L$ basis unless $\nu= 1/2$ or $3/2$. 
To make the calculation easier for tracing out the degrees of freedom in, say,
 the $L$ space later, we perform a further Bogoliubov transformation
 in each of $R$ and $L$ region. Apparently, this Bogoliubov transformation does 
not mix the operators in ${\cal H}_R$ space and those in ${\cal H}_L$ space. 
We introduce new operators $c_q=(c_R,c_L)$ that satisfy
\begin{eqnarray}
c_R = u\,b_R + v\,b_R^\dagger \,,\qquad
c_L = u^*\,b_L + v^*\,b_L^\dagger\,,
\label{bc}
\end{eqnarray}
to obtain
\begin{eqnarray}
|0\rangle_{\rm BD} = N_{\gamma_p}^{-1}
\exp\left(\gamma_p\,c_R^\dagger\,c_L^\dagger\,\right)|0\rangle_{R'}|0\rangle_{L'}\,.
\label{bogoliubov2}
\end{eqnarray}
Note that the condition $|u|^2-|v|^2=1$ is assumed
so that the new operators satisfy the commutation relation 
$[c_i,(c_j)^\dagger]=\delta_{ij}$. 
The normalization factor $N_{\gamma_p}$ is given by
\begin{eqnarray}
N_{\gamma_p}^2
=\left|\exp\left(\gamma_p\,c_R^\dagger\,c_L^\dagger\,\right)|0\rangle_{R'}|0\rangle_{L'}
\right|^2
=\frac{1}{1-|\gamma_p|^2}\,,
\label{norm2}
\end{eqnarray}
where $|\gamma_p|<1$ should be satisfied. The consistency relations from
 Eq.~(\ref{bogoliubov2}) 
($c_R|0\rangle_{\rm BD}=\gamma_p\,c_L^\dag|0\rangle_{\rm BD}$,
$c_L|0\rangle_{\rm BD}=\gamma_p\,c_R^\dag|0\rangle_{\rm BD}$) give
\begin{eqnarray}
\gamma_p=\frac{1}{2\zeta}
\left[-\omega^2+\zeta^2+1-\sqrt{\left(\omega^2-\zeta^2-1\right)^2-4\zeta^2}\,\right]\,,
\label{gammap}
\end{eqnarray}
where we defined $\omega\equiv m_{RR} = m_{LL}$ and $\zeta\equiv m_{RL}=m_{LR}$ in Eq.~(\ref{mij}). Note that a minus sign in front of the square root term is taken to make $\gamma_p$ converge. Putting the $\omega$ and $\zeta$ defined in Eq.~(\ref{mij}) 
into Eq.~(\ref{gammap}), we obtain 
\begin{eqnarray}
\gamma_p = i\frac{\sqrt{2}}{\sqrt{\cosh 2\pi p + \cos 2\pi \nu}
 + \sqrt{\cosh 2\pi p + \cos 2\pi \nu +2 }}\,.
\label{gammap2}
\end{eqnarray}
Note that $\gamma_p$ simplifies to $|\gamma_p|=e^{-\pi p}$
for $\nu=1/2$ (conformal) and $\nu=3/2$ (massless).
The $u$ and $v$ may be determined by inserting the above $\gamma_p$ into 
the consistency conditions.

\subsection{Reduced density matrix and entanglement entropy}

Given the density matrix in the diagonalized form,
it is straightforward to obtain the reduced density matrix.
From  Eqs.~(\ref{bogoliubov2}) and (\ref{norm2}), we obtain
the density matrix for each mode labeled by $p,\ell, m$ as
\begin{eqnarray}
\rho_R ={\rm Tr}_{L}\,|0\rangle_{\rm BD}\,{}_{\rm BD}\langle 0| 
=\left(1-|\gamma_p|^2\,\right)\sum_{n=0}^\infty 
|\gamma_p |^{2n}\,|n;p\ell m\rangle\langle n;p\ell m|\,,
\label{dm} 
\end{eqnarray}
where we defined $|n;p\ell m\rangle=1/\sqrt{n!}\,(c_R^\dagger)^n\,|0\rangle_{R'}$. 
In the conformal ($\nu=1/2$) and massless ($\nu=3/2$) cases,
the reduced density matrix reduces to a thermal state with temperature $T=H/(2\pi)$.

\begin{figure}[t]
\vspace{-2cm}
\includegraphics[height=7.8cm]{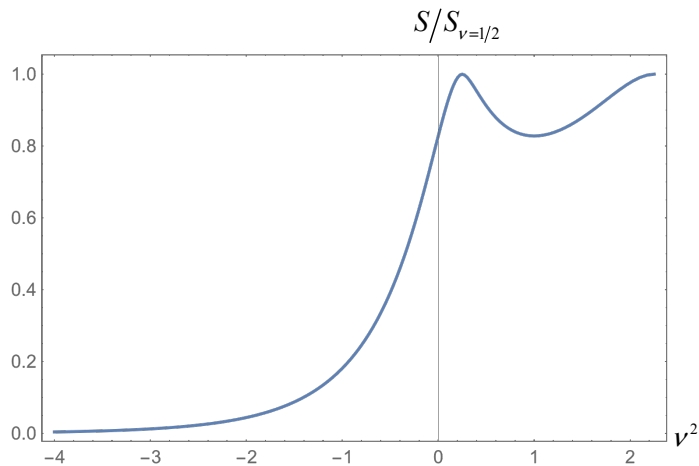}\centering
\vspace{-0.8cm}
\caption{Plot of the entanglement entropy normalized by the conformal
scalar case ($\nu=1/2$) as a function of $\nu^2$.}
\label{fig2}
\end{figure}

The entanglement entropy for each mode is given by
\begin{eqnarray}
S(p,\nu)=-{\rm Tr}\,\rho_R(p)\log_2\rho_R(p)
=-\log_2\left(1-|\gamma_p|^2\right)
-\frac{|\gamma_p|^2}{1-|\gamma_p|^2}\log_2|\gamma_p|^2\,.
\label{s}
\end{eqnarray}
Then the total entanglement entropy between two causally disconnected 
open regions is obtained by integrating over $p$ and a volume integral 
over the hyperboloid,
\begin{eqnarray}
S(\nu)=V_{H^3}^{\rm{reg}}\int_0^\infty\frac{dp\,p^2}{2\pi^2}S(p,\nu)
=\frac{1}{\pi}\int_0^\infty dp\,p^2S(p,\nu)\,,
\label{ints}
\end{eqnarray}
where $V_{H^3}^{\rm{reg}}=2\pi$ is the regularized volume of the 
hyperboloid~\cite{Maldacena:2012xp}.
The result is plotted in Figure~\ref{fig2}. We see that the entanglement 
is largest for small mass (positive $\nu^2$) and decays exponentially 
for large mass (negative $\nu^2$). The two peaks correspond 
to the massless ($\nu=3/2$) and conformal ($\nu=1/2$) cases.

\section{Effects of a bubble wall on the entanglement}

\begin{figure}[t]
\begin{center}
\vspace{-2cm}
\hspace{-1.5cm}
\begin{minipage}{8.0cm}
\includegraphics[height=8cm]{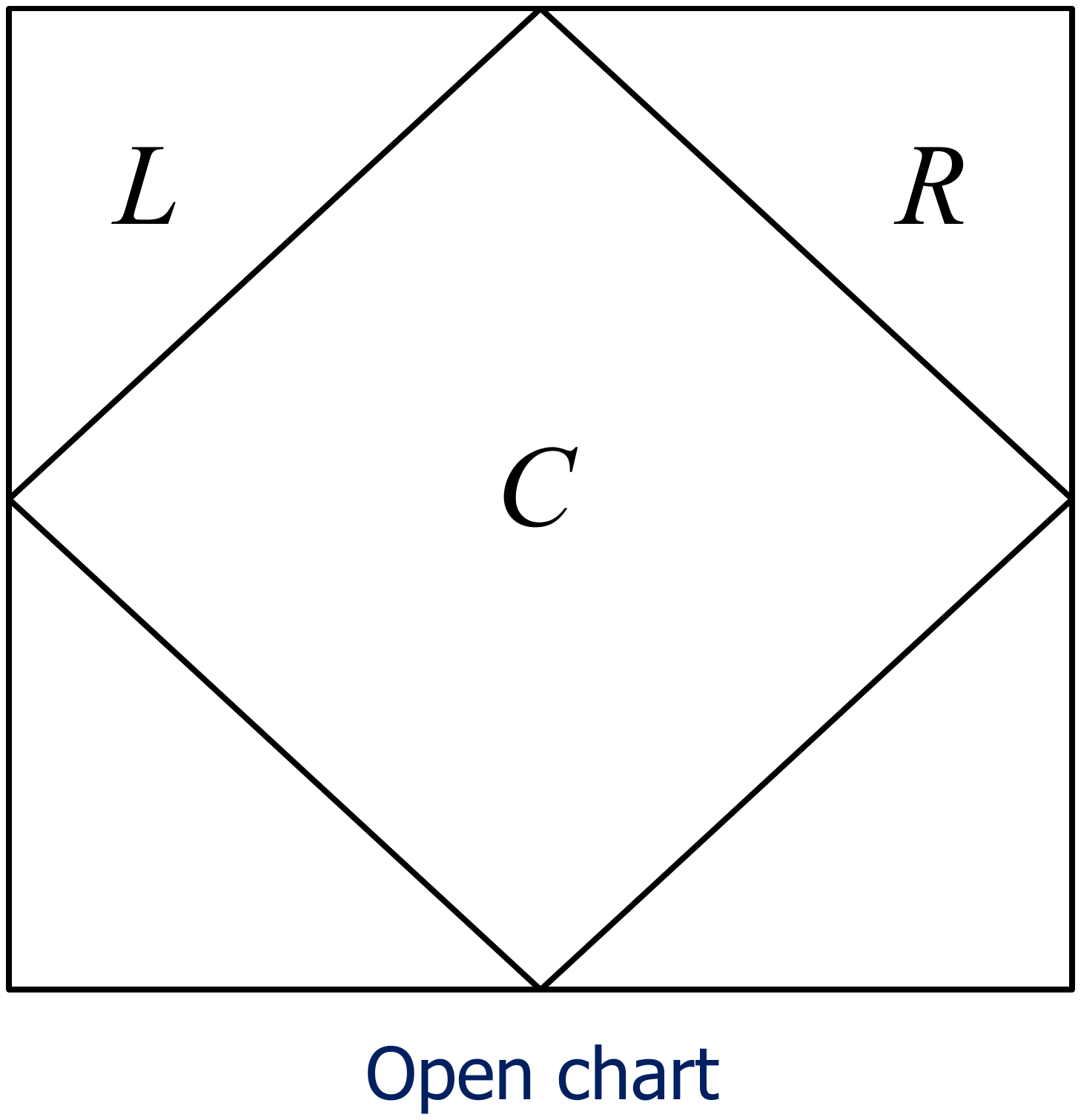}\centering
\end{minipage}
\begin{minipage}{8.0cm}
%\hspace{0.5cm}
\includegraphics[height=8cm]{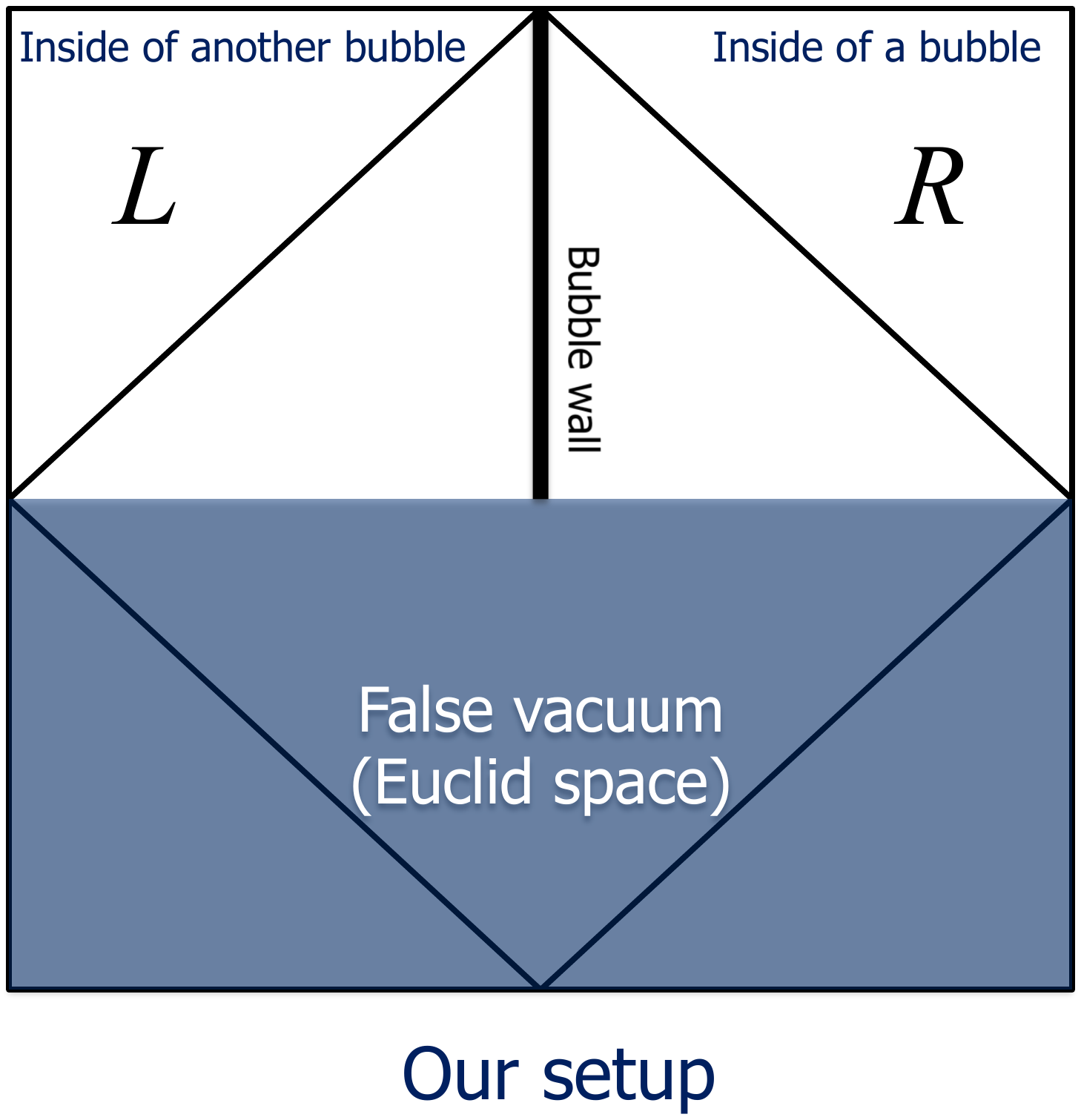}
\end{minipage}
\vspace{-0.8cm}
\caption{The Penrose diagrams of de Sitter space with
and without a delta function wall.
We assume that pair creation of identical vacuum bubbles 
through false vacuum decay, and that the bubbles are 
separated by an infinitesimally thin wall in region $C$.}
\label{fig3}
\end{center}
\end{figure}

Now we study the effect of a bubble wall on the entanglement.
 The Penrose diagram of our setup is depicted in Figure~\ref{fig3}. 
We consider the same action as Eq.~(\ref{action}) but now with
$m^2$ as a function of the background geometry which contains
a wall. In the case when the background geometry is given by
an instanton solution with $\sigma(\tau)$ being the scalar field 
configuration and $\phi$ being its fluctuations, $m^2$ will be given by
\begin{align}
m^2(\tau)=\frac{d^2V(\sigma)}{d\sigma^2}\,,
\end{align}
where $V$ is the potential of the $\sigma$ field,
and the $\tau$-dependence of $m^2$ is through its 
$\sigma$-dependence. In a realistic situation, $m^2$ would
be a smooth function of $\tau$, and is positive on both sides 
of the wall, but negative at the wall where the potential has a peak.
For simplicity, however, here we model the wall with a delta-function.

\subsection{Setup}

We consider the same action as Eq.~(\ref{action}) but now with
a delta-functional wall in region $C$ parameterized by $\Lambda$ according to,
\begin{eqnarray}
S=\int d^4 x\sqrt{-g}\left[\,-\frac{1}{2}\,g^{\mu\nu}
\partial_\mu\phi\,\partial_\nu \phi
-\frac{m^2-\Lambda\delta(t_C)}{2}\,\phi^2\,\right]\,.
\label{action2}
\end{eqnarray} 
where the metric is expressed as
\begin{eqnarray}
ds^2_C&=&H^{-2}\left[dt_C^2+\cos^2t_C\left(-dr_C^2+\cosh^2r_C\,d\Omega^2\right)\right]\,.
\end{eqnarray}
Note that the radial coordinate $t_C$ in the region $C$ 
coincides with $\tau$ of the instanton solution (see Eq.~(\ref{region:C}) below).
Note also that if we denote the width of the wall by $\Delta\tau_w$,
we have $\Lambda=|d^2V/d\sigma^2|H\Delta\tau_w$.

Setting the field $\phi$ as
\begin{eqnarray}
\phi = \frac{H}{\cos t_C}\chi_{p}(t_C)Y_{p\ell m}(r_C,\Omega)\,,
\end{eqnarray}
the solution of the mode function $\chi_p$ in the $C$ region is given 
by the associated Legendre function,
 $\chi_p\propto P^{\pm ip}_{\nu-\frac{1}{2}}(\sin t_C)$.

\subsection{Mode functions in the presence of a wall}

Now we want to pick up the positive frequency mode functions 
which are relevant for the pair creation of bubble universes
through false vacuum decay. Namely, those mode functions
that describe the Euclidean vacuum in the presence of a wall in region $C$.
They are obtained by requiring regularity in
the lower hemisphere of the Euclidean de Sitter space with the wall
when they are analytically continued to that 
region~\cite{Sasaki:1994yt,Yamamoto:1996qq}.

\subsubsection{The relation between the Lorentzian and the Euclidean coordinates}

The open chart is obtained by analytic continuation of the Euclidean sphere $S^4$.
The Lorentzian coordinates of the regions $L$, $R$ and $C$ are 
related to the Euclidean coordinates given in Eq.~(\ref{Emetric}) as
\begin{eqnarray}
\left\{
\begin{array}{l}
t_R=i\left(\tau-\frac{\pi}{2}\right)\hspace{2.2cm}\,,t_R\geq 0\\
r_R=i\rho\hspace{3.5cm}\,,r_R\geq 0
\end{array}
\right.
\label{region:R}
\end{eqnarray}
\begin{eqnarray}
\left\{
\begin{array}{l}
t_C=\tau\hspace{2.5cm}\,,-\frac{\pi}{2}\leq t_C\leq \frac{\pi}{2}\\
r_C=i\left(\rho-\frac{\pi}{2}\right)\hspace{1.0cm}\,,0\leq r_C\leq \infty
\end{array}
\right.
\label{region:C}
\end{eqnarray}
\begin{eqnarray}
\left\{
\begin{array}{l}
t_L=i\left(-\tau-\frac{\pi}{2}\right)\hspace{2cm}\,,t_L\geq 0\\
r_L=i\rho\hspace{3.6cm}\,,r_L\geq 0
\end{array}
\right.
\end{eqnarray}
For simplicity, we write $\sin t_C\equiv z_C$,
$\cosh t_R\equiv z_R$, and $\cosh t_L\equiv -z_L$ below. 
Then, the above relations give $z_C=z_R=-z_L$.

\subsubsection{Analytic continuation in the presence of the wall}

Let $\chi_p^R(z_R)=P^{ip}_{\nu-\frac{1}{2}}(z_R)$
and $\chi_p^L(z_L)=P^{ip}_{\nu-\frac{1}{2}}(z_L)$ ($z_L=-z_R$)
where
\begin{eqnarray}
P^\mu_\nu(z)=\frac{1}{\Gamma(1-\mu)}
\left(\frac{z+1}{z-1}\right)^{\frac{\mu}{2}}
F\left(-\nu,\nu+1,1-\mu;\frac{1-z}{2}\right)\,;
\quad z>1~\mbox{or}~z<-1\,.
\label{legendre1}
\end{eqnarray}

\vspace{6mm}
\noindent
$\bullet$ From $R$ ($R=\{z_R>1\}$) to $C^+$ ($C^+=\{0<z_C<1\}$):
\\

Analytic continuation is through $\Im z_R<0$.
This means that the argument of $z_R-1=z_C-1$ is $-\pi$. 
Hence $z_R-1=z_C-1=e^{-i\pi}(1-z_C)$.
Thus
\begin{eqnarray}
(1+z_C)=(1+z_R)\,,
\quad 
(1-z_C)=|1-z_R|e^{i\pi}=(z_R-1)e^{i\pi}\,,
\end{eqnarray}
which gives
\begin{eqnarray}
\left(\frac{1+z_R}{z_R-1}\right)^{i\frac{p}{2}}
=e^{-\frac{\pi}{2}p}\left(\frac{1+z_C}{1-z_C}\right)^{i\frac{p}{2}}\,,
\end{eqnarray}
when analytically continued from $z_R>1$ to $z_R=z_C<1$.
This means
\begin{eqnarray}
\chi_p^R(z_C)=e^{-\frac{\pi}{2}p}\,\tilde{P}^{ip}_{\nu-\frac{1}{2}}(z_C)
\,,
\label{RtoC}
\end{eqnarray}
for $z_R=z_C<1$, where $\tilde{P}^\mu_\nu(x)$ for 
$-1<x<1$ is defined as
\begin{eqnarray}
\tilde{P}^\mu_\nu(x)=\frac{1}{\Gamma(1-\mu)}
\left(\frac{1+x}{1-x}\right)^{\frac{\mu}{2}}
F\left(-\nu,\nu+1,1-\mu;\frac{1-x}{2}\right)\,.
\label{legendre2}
\end{eqnarray}

\vspace{6mm}

\noindent
$\bullet$ From $C^+$ to $C^{-}$ ($C^-=\{-1<z_C<0\}$):

Assuming there is a delta-functional wall of height $\Lambda$
at $z_C=0$, $\chi_p^R(z_C)$ is deformed to
\begin{eqnarray}
\chi_p^R(z_C)=e^{\frac{\pi}{2}p}
\left(A_p e^{-\pi p}\tilde{P}^{ip}_{\nu-\frac{1}{2}}(z_C)
+B_p e^{\pi p}\tilde{P}^{-ip}_{\nu-\frac{1}{2}}(z_C)\right)\,,
\label{chiCminus}
\end{eqnarray}
in $C^-$, where $A_p$ and $B_p$ are given by
\begin{eqnarray}
A_p&=&1+\frac{\pi}{2i\sinh\pi p}\frac{\Lambda}{H^2}
|\tilde{P}^{ip}_{\nu-\frac{1}{2}}(0)|^2\,,
\\
B_p&=&-\frac{\pi}{2i\sinh\pi p}\frac{\Lambda}{H^2}
e^{-2\pi p}\left(\tilde{P}^{ip}_{\nu-\frac{1}{2}}(0)\right)^2\,.
\end{eqnarray}
Note that in the absence of a wall ($\Lambda=0$), we have $A_p=1$ and $B_p=0$.

%Here we note that
%\begin{align}
%\overline{A_p}=A_{-p}\,,\quad
%\overline{B_p}=e^{-4\pi p}B_{-p}\,.
%\label{ABprop}
%\end{align}

\vspace{6mm}

\noindent
$\bullet$ From $L$ ($L=\{z_L<-1\}$) to $C^-$ ($C^-=\{-1<z_C<0\}$):

Now we express the above solution in terms of $\chi^L_p$.
To do this, we first introduce $\hat{z}_C=-z_C$ and analytically
continue  $\chi^L_p$ from $L$ to $C^-$ where $0<\hat{z}_C<1$.
Exactly the same as the analytic continuation from $R$ to $C^+$,
we have
\begin{eqnarray}
\chi_p^L(\hat{z}_C)=e^{-\frac{\pi}{2}p}\tilde{P}^{ip}_{\nu-\frac{1}{2}}(\hat{z}_C)
\,,\label{LtoC}
\end{eqnarray}
for $z_L=\hat{z}_C<1$.

\vspace{6mm}

\noindent
$\bullet$ matching $\chi^R$ with $\chi^L$:

We now express $\chi^R_p$ in terms of $\chi^L_p$
and $\chi^L_{-p}$. To do this, we express $P^{\pm ip}_{\nu-\frac{1}{2}}(z_C)
=P^{\pm ip}_{\nu-\frac{1}{2}}(-\hat{z}_C)$  
in terms of $P^{\pm ip}_{\nu-\frac{1}{2}}(\hat{z}_C)$,
which can be achieved by using the transformation formulas for the 
hypergeometric functions in Appendix~\ref{app:b}. We find
\begin{eqnarray}
P^{ip}_{\nu-\frac{1}{2}}(-\hat{z}_C)
=C_p\,P^{ip}_{\nu-\frac{1}{2}}(\hat{z}_C)+D_p\, P^{-ip}_{\nu-\frac{1}{2}}(\hat{z}_C)\,,
\label{Ptrans}
\end{eqnarray}
where
\begin{eqnarray}
C_p=\frac{\cos\pi\nu}{i\sinh\pi p}\,,
\qquad
D_p=-e^{-2\pi p}\,
\frac{\cos\left(\nu+ip\right)\pi}{i\sinh\pi p}
\frac{\Gamma\left(\frac{1}{2}+\nu+ip\right)}{\Gamma\left(\frac{1}{2}+\nu-ip\right)}\,.
\end{eqnarray}
Using (\ref{Ptrans}), $\chi^R_p$ is expressed as\footnote{In the language 
of~\cite{Yamamoto:1996qq}, we have
\begin{eqnarray}
\alpha_p=e^{\pi p}\left(A_pD_{p}+B_pC_{-p}\right)\,,
\quad
\beta_p=A_pC_{p}+B_pD_{-p}\,.
\label{alphabeta}
\end{eqnarray}
We can check the symmetry,
$\alpha^*_p=e^{-2\pi p}\alpha_{-p}\,,~
\beta^*_p=\beta_{-p}\,.$}
\begin{eqnarray}
\chi^R_p(z_C)
&=&e^{\frac{\pi}{2}}\left[A_p\,P^{ip}_{\nu-\frac{1}{2}}(-\hat{z}_C)
+B_p\,P^{-ip}_{\nu-\frac{1}{2}}(-\hat{z}_C)\right]
\nonumber\\
&=&\left(A_pC_p+B_pD_{-p}\right)
\chi^L_{p}(\hat{z}_C)
+e^{\pi p}\left(A_pD_p+B_pC_{-p}\right)\chi^L_{-p}(\hat{z}_C)\,.
\end{eqnarray}
Notice that $P^{-ip}_{\nu-\frac{1}{2}}(-\hat{z}_C)$ is not complex conjugate of 
$P^{ip}_{\nu-\frac{1}{2}}(-\hat{z}_C)$.

\subsubsection{The Euclidean vacuum in the presence of the wall}

Finally, the positive frequency mode functions for the
Euclidean vacuum in the presence of the bubble wall are found to be
\begin{eqnarray}
\chi_p^R(z)&=&\frac{1}{N_w}\left\{
\begin{array}{l}
P_{\nu-\frac{1}{2}}^{ip}(z_R)
\,,\\\\
\left(A_pC_p+B_pD_{-p}\right)P_{\nu-\frac{1}{2}}^{ip}(z_L)
+e^{\pi p}\left(A_pD_p+B_pC_{-p}\right)P_{\nu-\frac{1}{2}}^{-ip}(z_L)\,,
\label{bwsolutions1}
\end{array}
\right.\\\nonumber\\
\chi_p^L(z)&=&\frac{1}{N_w}\left\{
\begin{array}{l}
\left(A_pC_p+B_pD_{-p}\right)P_{\nu-\frac{1}{2}}^{ip}(z_R)
+e^{\pi p}\left(A_pD_p+B_pC_{-p}\right)P_{\nu-\frac{1}{2}}^{-ip}(z_R)
\,,\\\\
P_{\nu-\frac{1}{2}}^{ip}(z_L)
\,,
\label{bwsolutions2}
\end{array}
\right.
\end{eqnarray}
where the Klein-Gordon normalization for the above solutions is
\begin{eqnarray}
N_w^2=\tilde{N_p}^2\left(1+|f_p|^2-|g_p|^2\right)\,,
\end{eqnarray}
where $\tilde{N}_p$ is defined in Eq.~(\ref{varphi}) and we have defined
\begin{eqnarray}
f_p=A_pC_p+B_pD_{-p}\,,\qquad g_p=e^{\pi p}\left(A_pD_p+B_pC_{-p}\right)\,.
\label{fg}
\end{eqnarray}
Note that in the absence of the wall ($\Lambda=0$), 
we have $f_p=C_p$ and $g_p=e^{\pi p}D_p$.

\subsection{Bogoliubov transformations and entangled states}

We perform the same Bogoliubov transformation in Eq.~(\ref{fo}) that mixes 
the operators in the Hilbert spaces ${\cal H}_R$ and ${\cal H}_L$.
The derivation of the symmetric matrix $m_{ij}$ is given in 
Appendix~\ref{app:c}. The components of the $m_{ij}$ in Eq.~(\ref{mij})
 is now expressed as
\begin{align}
\omega&=&-F\left[
\left(f_p+f_p^*\right)\left(1-\frac{|g_p|^2}{1-f_p^{*2}}\right)
-\left(1+|f_p|^2-|g_p|^2\right)\left(f_p-\frac{f_p^*|g_p|^2}{1-f_p^{*2}}\right)
\right]\,,\qquad
\label{omega}
\\
\zeta&=&-F\left[
\left(f_p+f_p^*\right)\left(f_p-\frac{f_p^*|g_p|^2}{1-f_p^{*2}}\right)
-\left(1+|f_p|^2-|g_p|^2\right)\left(1-\frac{|g_p|^2}{1-f_p^{*2}}\right)
\right]\,,\qquad
\label{zeta}
\end{align}
where
\begin{eqnarray}
F=\frac{g_p^*}{1-f_p^{*2}}\frac{1}{E}\,,
\qquad
E=\left(1-\frac{|g_p|^2}{1-f_p^{*2}}\right)^2
-\left(f_p-\frac{f_p^*|g_p|^2}{1-f_p^{*2}}\right)^2\,.
\end{eqnarray}

If there is no wall, $\omega$ is real ($\omega=\omega^*$) and $\zeta$ is pure imaginary ($\zeta=-\zeta^*$) for positive $\nu^2$, then the second Bogoliubov transformation was simplified as in Eq.~(\ref{bc}). In the presence of the wall, however, Eqs~(\ref{omega}) and (\ref{zeta}) are not real and pure imaginary respectively for positive $\nu^2$. In this case,
we need to perform the Bogoliubov transformation of the form
\begin{eqnarray}
c_R = u\,b_R + v\,b_R^\dagger \,,\qquad
c_L = \bar{u}\,b_L + \bar{v}\,b_L^\dagger\,,
\end{eqnarray}
to get the relation Eq.~(\ref{bogoliubov2}). Note that $|u|^2-|v|^2=1$ and $|\bar{u}|^2-|\bar{v}|^2=1$ are assumed. Then the consistency relations ($c_R|0\rangle_{\rm BD}=\gamma_p\,c_L^\dag|0\rangle_{\rm BD}$,
$c_L|0\rangle_{\rm BD}=\gamma_p\,c_R^\dag|0\rangle_{\rm BD}$) give the system of four homogeneous equations
\begin{eqnarray}
&&\omega\,u+v-\gamma_p\,\zeta\,\bar{v}^*=0\,,\qquad
\zeta\,u-\gamma_p\,\bar{u}^*-\gamma_p\,\omega\,\bar{v}^*=0\,,\nonumber\\
&&\omega\,\bar{u}+\bar{v}-\gamma_p\,\zeta\,v^*=0\,,\qquad
\zeta\,\bar{u}-\gamma_p\,u^*-\gamma_p\,\omega\,v^*=0\,.
\end{eqnarray}

\begin{figure}[t]
\begin{center}
\vspace{-2cm}
\hspace{-1.2cm}
\begin{minipage}{8cm}
\includegraphics[height=6.8cm]{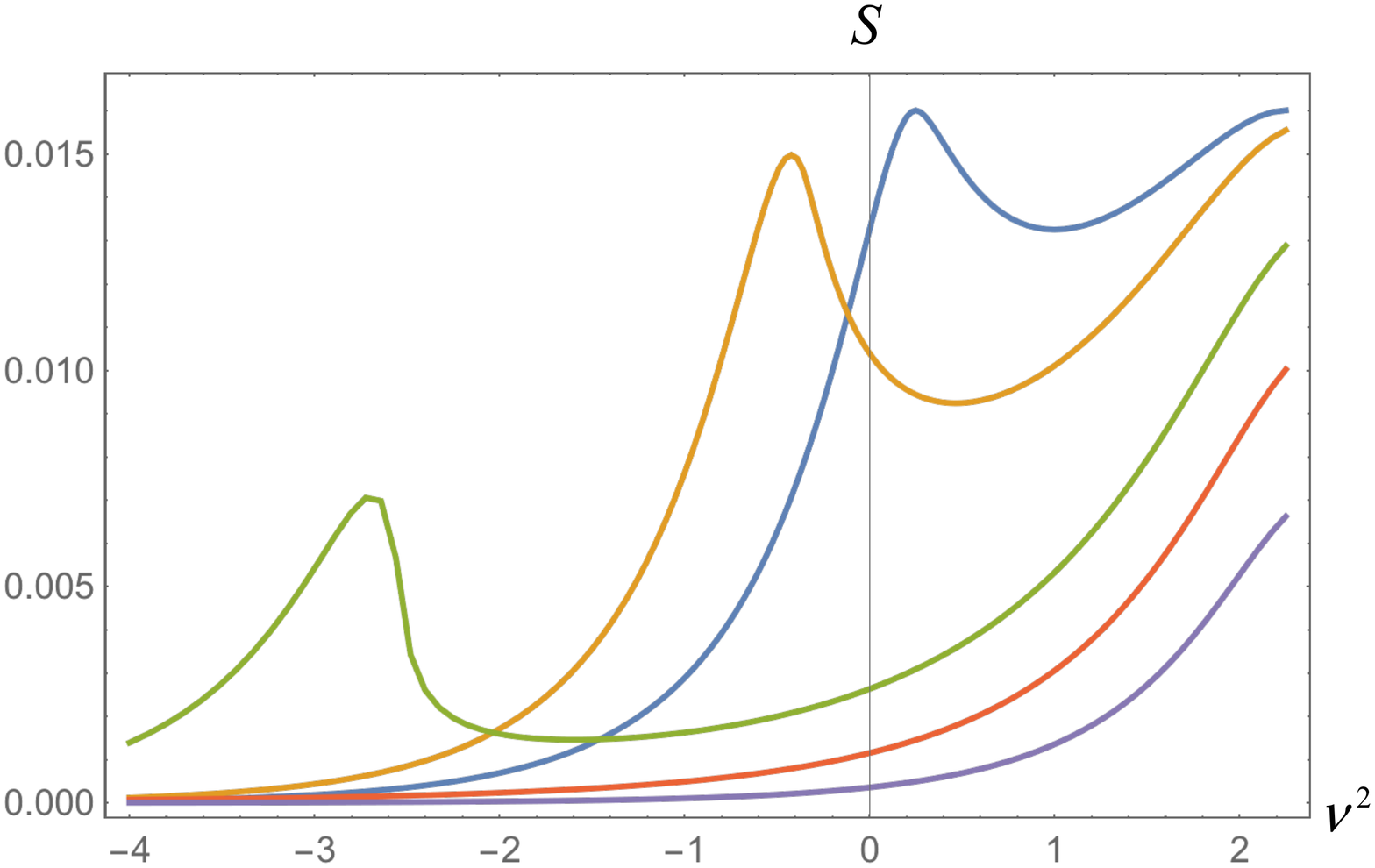}\centering
\end{minipage}
\begin{minipage}{8cm}
\hspace{0.2cm}
\includegraphics[height=6.8cm]{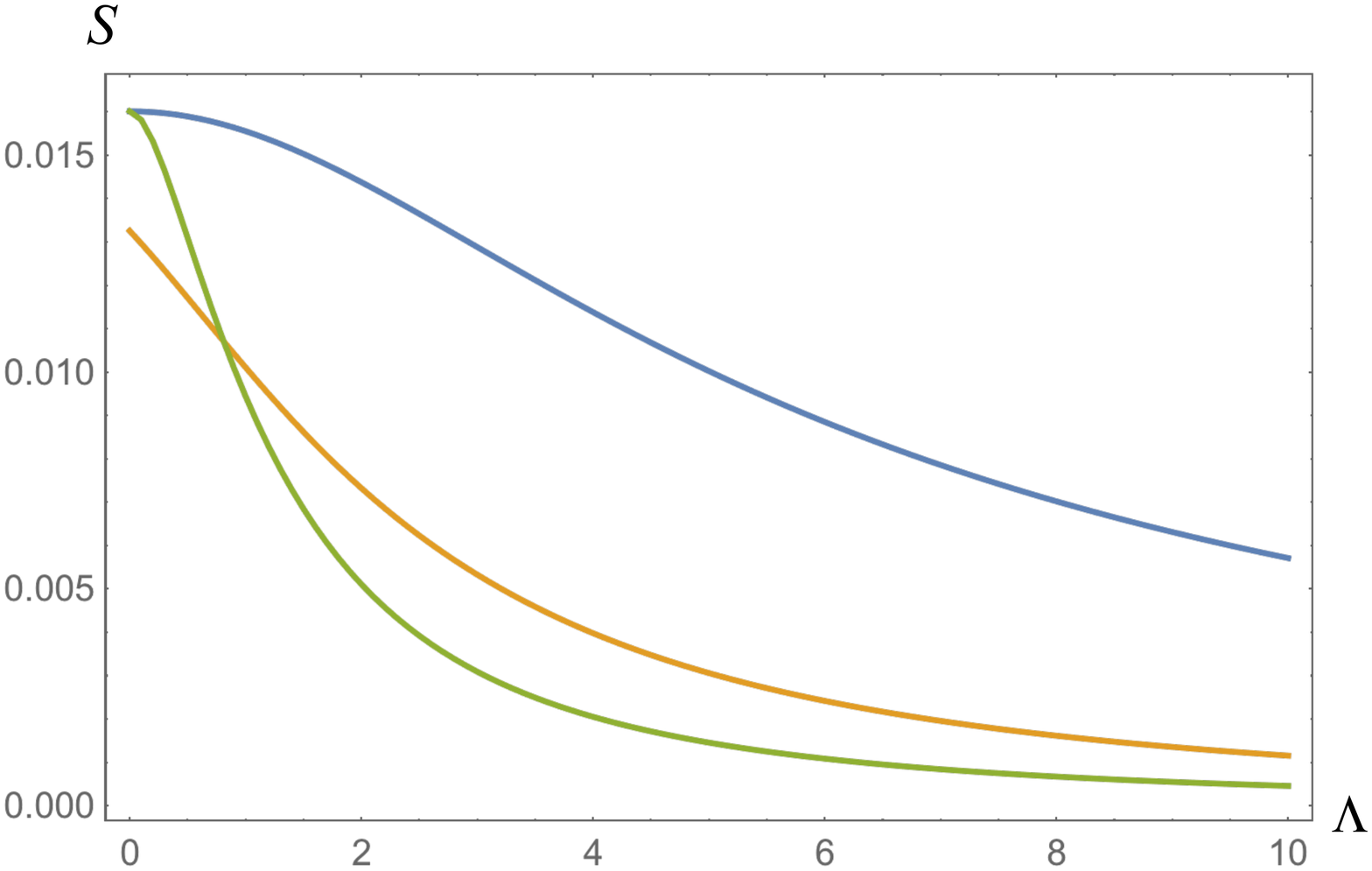}
\end{minipage}
\vspace{-0.5cm}
\caption{The left panel shows plots of the entanglement entropy versus $\nu^2$
for several values of $\Lambda$. Running from top to bottom on the right side of the panel: $\Lambda=0$ (blue), $\Lambda/H^2=1$ (orange), $\Lambda/H^2=3$ (green), $\Lambda/H^2=5$ (red)  and $\Lambda/H^2=8$ (purple).
The right panel shows the $\Lambda$ dependence of the entanglement
entropy, where the horizontal axis
is in units of $H=1$. Again running from top to bottom along the right, we show 
the massless case ($\nu=3/2$, blue), $\nu=1$ (orange) and the conformal case ($\nu=1/2$, green).}
\label{fig4}
\end{center}
\end{figure}

In order to have a non-trivial solution in the above system of equations, $\gamma_p$ must be~\cite{Kanno:2014lma}
\begin{eqnarray}
|\gamma_p|^2&=&\frac{1}{2|\zeta|^2}\left[
-\omega^2\zeta^{*2}-\omega^{*2}\zeta^2+|\omega|^4-2|\omega|^2+1+|\zeta|^4\right.\nonumber\\
&&\qquad\quad\left.-\sqrt{\left(\omega^2\zeta^{*2}+\omega^{*2}\zeta^2-|\omega|^4+2|\omega|^2-1-|\zeta|^4\right)^2-4|\zeta|^4}\,
\right]\,,
\label{gammap3}
\end{eqnarray}
where we took a minus sign in front of the square root term to reduce Eq.~(\ref{gammap}) when there is no wall.
Then putting Eqs.~(\ref{omega}) and (\ref{zeta}) into 
Eq.~(\ref{gammap3}), we can calculate the entanglement entropy for each 
mode in Eq.~(\ref{s}). The resulting total entanglement entropy, Eq.~(\ref{ints}),
is plotted in Figure~\ref{fig4}.

\subsection{Entanglement entropy}

From the left panel in Figure~\ref{fig4}, we see that the entanglement entropy decreases as the effect of the wall increases for small mass (positive $\nu^2$). 
For large mass (negative $\nu^2$), the entanglement entropy decays 
exponentially in the absence of the wall ($\Lambda=0$). In the presence of the wall, the peak at the conformally coupled scalar ($\nu=1/2$) shifts to the 
left and eventually disappears as the effect of the wall increases.
 The right panel also shows that the peak of the entanglement entropy
 corresponding to the massless case and the conformally coupled scalar
 is an identical value in the absence of the wall ($\Lambda=0$). 
However, as the effect of the wall becomes large, the entanglement entropy 
in the case of conformally coupled scalar decays faster than that of massless case.

\subsection{Logarithmic negativity}
\label{ln}

In order to characterize the entanglement of a quantum state, there have been many entanglement measures proposed. The logarithmic negativity is one such measure of quantum entanglement. This measure is derived from the positive partial transpose criterion for separability~\cite{Horodecki:2009zz}. The idea of it is to characterize an entangled state as a state that is not separable. In this subsection, we compute the entanglement of our model with the logarithmic negativity.

\begin{figure}[t]
\begin{center}
\vspace{-2cm}
\hspace{-1.2cm}
\begin{minipage}{8.0cm}
\includegraphics[height=6.8cm]{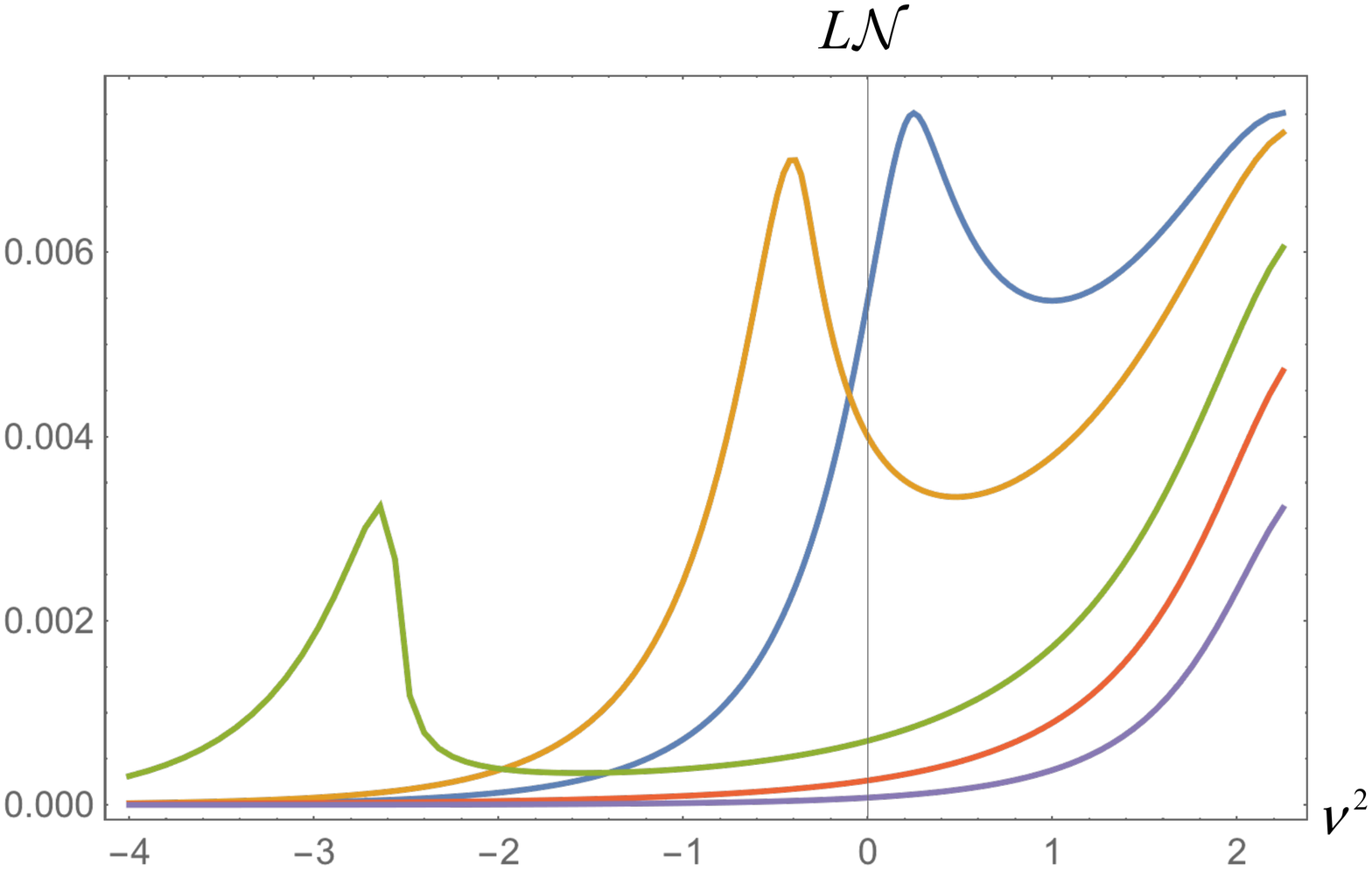}\centering
\end{minipage}
\begin{minipage}{8.0cm}
\hspace{0.2cm}
\includegraphics[height=6.8cm]{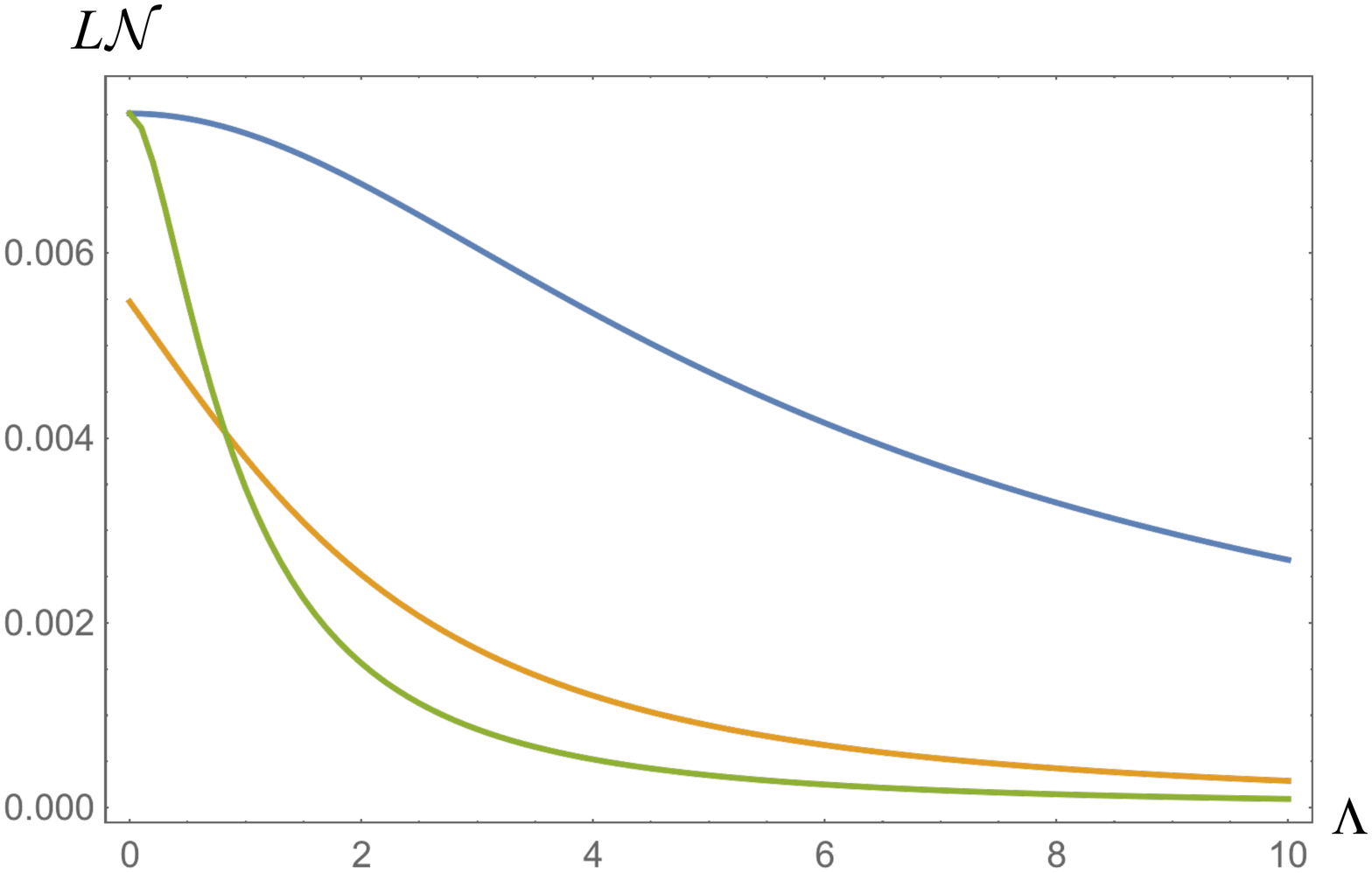}
\end{minipage}
\vspace{-0.5cm}
\caption{The left panel shows plots of the logarithmic negativity versus $\nu^2$. We set $H=1$.
Running from top to bottom on the right side of the panel: $\Lambda=0$ (blue), $\Lambda/H^2=1$ (orange), $\Lambda/H^2=3$ (green), $\Lambda/H^2=5$ (red)  and $\Lambda/H^2=8$ (purple).
The right panel shows the $\Lambda$ dependence of the logarithmic negativity, where the horizontal axis is in units of $H=1$. Again running from top to bottom along the right, we show 
the massless case ($\nu=3/2$, blue), $\nu=1$ (orange) and the conformal case ($\nu=1/2$, green).}

\label{fig5}
\end{center}
\end{figure}

The second Bogoliubov transformation Eq.~(\ref{bogoliubov2}) is rewritten as
\begin{eqnarray}
|0\rangle_{\rm BD} = N_{\gamma_p}^{-1}
\sum_{n=0}^\infty\gamma_p^n\,|n;p\ell m\rangle_{R'}|n;p\ell m\rangle_{L'}\,,
\label{bogoliubov3}
\end{eqnarray}
where the states $|n;p\ell m\rangle_{R'}$ and $|n;p\ell m\rangle_{L'}$ are $n$ particle excitation states in $R'$ and $L'$ spaces. For a pure state, any state has a Schmidt decomposition expressed as
\begin{eqnarray}
  |\psi\rangle=\sum_i\sqrt{\lambda_i}\,|i\rangle_A\otimes|i\rangle_B\,,
  \label{lnschmidt}
\end{eqnarray}
where $\lambda_i$ is the probability to observe the $i$th state and satisfies $\sum_i\lambda_i=1$. By using the eigenvalues, the logarithmic negativity is expressed as
\begin{eqnarray}
L{\cal N}=2\log_2\left(\sum_i\sqrt{\lambda_i}\right)\,.
\end{eqnarray}
Thus, if we compare Eq.~(\ref{bogoliubov3}) with the Schmidt decomposition, we can read off the corresponding eigenvalues
\begin{eqnarray}
\sqrt{\lambda_i}=N_{\gamma_p}^{-1}|\gamma_p|^n\,,
\end{eqnarray}
and the logarithmic negativity for each mode is calculated as~\cite{Kanno:2014bma}
\begin{eqnarray}
L{\cal N}(p,\,\nu)=2\log_2\left(\sum_i N_{\gamma_p}^{-1}|\gamma_p|^n\right)=\log_2\frac{1+|\gamma_p|^2}{1-|\gamma_p|^2}\,.
\end{eqnarray}
Then the logarithmic negativity is obtained by integrating over $p$ and a volume integral over the hyperboloid,
\begin{eqnarray}
L{\cal N}(\nu)=\frac{1}{\pi}\int_0^\infty dp\,p^2 L{\cal N}(p,\,\nu)\,.
\end{eqnarray}
The result is plotted in Figure~\ref{fig5}. We find that the qualitative features are the same as the result of entanglement entropy.

\section{Summary and discussion}

We have studied the effect of a bubble wall on the entanglement entropy of 
a free massive scalar field between two causally disconnected open charts in 
de Sitter space. We assume there is a delta-functional wall between them parameterized by our wall parameter $\Lambda$.
This may be regarded as a model describing the pair creation of
identical bubble universes separated by a bubble wall. 
To analyze the system, we first derived the Euclidean vacuum mode functions 
of the scalar field in the presence of the wall in the coordinates that
 respect the open charts. 
We then gave the Bogoliubov transformation between the Euclidean vacuum and
 the open chart vacua that makes the reduced density matrix diagonal. 
We derived the reduced density matrix in one of the open charts ($R$ space) 
after tracing out the other ($L$ space). We then computed the entanglement entropy of the scalar field by using the reduced density matrix and compared the result with the case of no bubble wall. We found that larger values of parameter $\Lambda$ correspond to less entanglement. We also computed a different measure of entanglement called logarithmic negativity. The qualitative features were found to be the same as the result of entanglement entropy.

In the limit of small entanglement entropy the BD quantum state approaches a product of ground state wavefunctions for each of the charts. Our results thus show that for large $\Lambda$ the dynamics of bubble formation select this product state and ensure its stability under evolution.  These are the features identified in the literature\footnote{See for example~\cite{Zurek:1981xq,Zurek:1992mv,Zurek:1992aa1,Anglin:1995pg} and also~\cite{Zurek:2003zz}, section IV-D, for a nice review.} to correspond to the selection of special “pointer states” via the decoherence process. Our results thus may be regarded as evidence of decoherence of bubble universes from (and by) each other.

We also note that in discussions of the black hole firewall problem~\cite{Braunstein:2009my,Almheiri:2012rt} it is argued that the absence of entanglement implies the existence of a firewall. We are intrigued by a certain parallel, in a kind of reverse engineered way, with our results: We show a particular example of how the presence of a wall can reduce entanglement.

\section*{Acknowledgments}
This work was supported in part by the MEXT KAKENHI Nos.~15H05888 and 15K21733.
SK was supported by IKERBASQUE, the Basque Foundation 
for Science and the Basque Government (IT-979-16),  
and Spanish Ministry MINECO  (FPA2015-64041-C2-1P). AA was supported by a grant from UC Davis, and thanks A. Arrasmith for helpful conversations.

\appendix

\section{Supercurvature mode}
\label{app:a}

On the de Sitter background, there exists a supercurvature mode 
in the open chart if the mass-squared is in the range 
$0<m^2/H^2<2$~\cite{Sasaki:1994yt}. The supercurvature mode has an 
imaginary eigenvalue, $p=ik$ where $0<k<1$.
Therefore this may be regarded as a bound-state mode in the spectrum.

The role of supercurvature modes in the quantum entanglement is not known.
In \cite{Maldacena:2012xp}, it is conjectured that they do not 
contribute to the entanglement. Here we consider the effect of the
presence of a bubble wall on the supercurvature mode by focusing
on the simplest case of $m^2/H^2=2$, ie, the conformal scalar case.
In this case, if there is no wall, there is no supercurvature mode.
We see below that a supercurvature mode appears when there is a wall.

Let us first write down the equation for the mode function $\chi_p(t_C)$
in region $C$,
\begin{align}
\left[\frac{d^2}{dt_C^2}+\frac{da(t_C)}{a(t_C)dt_C}\frac{d}{dt_C}
+2-M^2(t_C)+\frac{p^2}{H^2a^2(t_C)}\right]\chi_p=0\,,
\label{modeCteq}
\end{align}
where $a(t_C)=H^{-1}\cos t_C$ and 
\begin{align}
M^2=\frac{m^2-\Lambda\delta(t_C)}{H^2}\,.
\end{align}
By using the conformal coordinate $d\xi=dt_C/a(t_C)$,
we have $a=(H\cosh\xi)^{-1}$ and Eq.~(\ref{modeCteq}) is re-expressed as
\begin{align}
\left[-\frac{d^2}{d\xi^2}+\frac{M^2-2}{\cosh^2\xi}-p^2\right]\chi_p=0\,.
\label{modeCxieq}
\end{align}

We set $p^2=-k^2$ $(k>0)$, since we consider a supercurvature mode. 
For $M^2=2$, Eq.~(\ref{modeCxieq}) gives the solution,
\begin{align}
\chi_k\propto e^{\pm k\xi}\qquad\mbox{for}\quad\xi\neq0\,.
\end{align}
For either sign, the solution is singular at $\xi\to\mp\infty$
if there were no wall.
However, the presence of a wall allows the solution to be
\begin{align}
\chi_k\propto
\left\{
\begin{array}{ll}
e^{-k\xi}\quad&\mbox{for}~\xi>0\,,
\\
e^{k\xi} \quad&\mbox{for}~\xi<0\,.
\end{array}
\right.
\end{align}
The matching condition at $\xi=0$ gives
\begin{align}
\left[-\frac{d}{d\xi}\chi_p\right]^{+}_{-}=\frac{\Lambda}{H^2}\chi_p\,,
\end{align}
which can be readily solved to obtain
\begin{align}
k=\frac{\Lambda}{2H^2}\,.
\end{align}
Thus the supercurvature mode exists for any value of $\Lambda>0$,
and the eigenvalue $k$ can be arbitrarily large unlike the case
of the pure de Sitter background.

To complete the analysis, let us compute the normalization factor 
of the supercurvature mode.
Setting $\chi_k=N_k^{-1}e^{\pm k\xi}$ for $\xi\lessgtr 0$, we have
\begin{align}
1=\int_{-\infty}^\infty d\xi |\chi_k|^2
=\frac{2}{N_k^{2}}\int_{0}^\infty d\xi e^{-2k\xi}=\frac{1}{N_k^2k}\,.
\end{align}
Thus we obtain a very simple result,
\begin{align}
N_k=\frac{1}{\sqrt{k}}\,.
\end{align}
Thus the larger the eigenvalue $k$, the smaller the normalization
factor becomes, implying that its contribution to the spectrum
of the vacuum fluctuations in each open chart becomes more and more important~\cite{Yamamoto:1996qq}.

\section{Transformation formulas}
\label{app:b}

\begin{eqnarray}
F\left(\alpha,\beta,\gamma;z\right)&=&\frac{\Gamma(\gamma)\Gamma(\alpha+\beta-\gamma)}{\Gamma(\alpha)\Gamma(\beta)}\left(1-z\right)^{\gamma-\alpha-\beta}F\left(\gamma-\alpha\,,\gamma-\beta\,,\gamma-\alpha-\beta+1\,;1-z\right)\nonumber\\
&&+\frac{\Gamma(\gamma)\Gamma(\gamma-\alpha+\beta)}{\Gamma(\gamma-\alpha)\Gamma(\gamma-\beta)}
F\left(\alpha\,,\beta\,,\alpha+\beta-\gamma+1\,;1-z\right)\,,
\end{eqnarray} 
and
\begin{eqnarray}
F\left(\alpha,\beta,\gamma;z\right)=\left(1-z\right)^{\gamma-\alpha-\beta}
F\left(\gamma-\alpha\,,\gamma-\beta\,,\gamma\,;z\right)\,.
\end{eqnarray}

\section{Bogoliubov coefficients}
\label{app:c}

From Eq.~(\ref{fo}), the relation between the operators $a_I$ and $b_J$ is given by
\begin{eqnarray}
b_J=a_I\left(M\right)^I{}_J\,,\qquad b_J=\left(b_q\,,b_q^\dag\right)\,,\qquad
a_I=\left(a_\sigma\,,a_\sigma^\dag\right)\,,
\label{rel:ba}
\end{eqnarray}
where the capital indices $\left(I\,,J\right)$ run from $1$ to $4$, the subscripts $q,\sigma=\left(R\,,L\right)$ and $M$ is a $4\times 4$ matrix
\begin{eqnarray}
M^I{}_J=\left(
\begin{array}{cc}
\alpha^\sigma{}_q & \beta^\sigma{}_q\\
\beta^{\sigma*}{}_{\!\!\!q} & \alpha^{\sigma*}{}_{\!\!\!q} \\
\end{array}\right)\,,
\end{eqnarray}
and $\alpha$ and $\beta$ are $2\times 2$ matrices and consist of $f_p$ and $g_p$ in Eq.~(\ref{fg}) such as
\begin{eqnarray}
\alpha^\sigma{}_q=\frac{\tilde{N}_p}{N_w}\left(
\begin{array}{cc}
1 & f_p \\
f_p & 1 \\
\end{array}\right)\,,\qquad
\beta^\sigma{}_q=\frac{\tilde{N}_p}{N_w}\left(
\begin{array}{cc}
0 & g_p \\
g_p & 0 \\
\end{array}\right)\,.
\label{alphabeta}
\end{eqnarray}
The relation (\ref{rel:ba}) is rewritten as
\begin{eqnarray}
a_J=b_I\left(M^{-1}\right)^I{}_J\,,\qquad
\left(M^{-1}\right)^I{}_J=\left(
\begin{array}{ll}
\xi_{q\sigma} & \delta_{q\sigma} \vspace{3mm}\\
\delta_{q\sigma}^* & \xi_{q\sigma}^* \\
\end{array}\right)\,,\qquad
\left\{
\begin{array}{l}
\xi=
\left(\alpha-\beta\,\alpha^{*\,-1}\beta^*\right)^{-1}\,,\vspace{3mm}\\
\delta=-\alpha^{-1}\beta\,\xi^*\,.
\end{array}
\right.
\label{xidelta1}
\end{eqnarray}
By using Eq.~(\ref{alphabeta}), we find the above matrices $\xi$ and $\delta$ are expressed respectively as
\begin{eqnarray}
\xi&=&\frac{N_w}{\tilde{N}_p}\frac{1}{E}
\left(
\begin{array}{cc}
1-\frac{|g_p|^2}{1-f_p^{*2}} & -f_p-\frac{f_p^*|g_p|^2}{1-f_p^{*2}}\\
-f_p-\frac{f_p^*|g_p|^2}{1-f_p^{*2}} & 1-\frac{|g_p|^2}{1-f_p^{*2}} \\
\end{array}\right)\,,\\
\delta&=&\frac{N_w}{\tilde{N}_p}\frac{1}{E^*}\frac{g_p}{1-f_p^2}
\left(
\begin{array}{cc}
f_p+f_p^* & -1-|f_p|^2+|g_p|^2\\
-1-|f_p|^2+|g_p|^2 & f_p+f_p^* \\
\end{array}\right)\,.
\end{eqnarray}
If we apply $a_\sigma$ to Eq.~(\ref{bogoliubov1}), then we have
\begin{eqnarray}
0=a_\sigma\,|0\rangle_{\rm BD}\Longrightarrow
m_{ij}=-\left(\delta^*\xi^{-1}\right)_{ij}\,,
\end{eqnarray}
and $m_{ij}$ is found to be
\begin{eqnarray}
m_{ij}=\left(
\begin{array}{cc}
\omega & \zeta\\
\zeta & \omega \\
\end{array}\right)\,,
\end{eqnarray}
where $\omega$ and $\zeta$ are given in Eqs.~(\ref{omega}) and (\ref{zeta}).

\end{document}